# Dissipative Soliton Fiber Lasers with Higher-Order Nonlinearity, Multiphoton Absorption and Emission, and Random Dispersion

GURKIRPAL SINGH PARMAR,[1] SOUMENDU JANA,[1,*] AND BORIS A. MALOMED[2,3]

[1] School of Physics and Materials Science, Thapar University, Patiala-147004, India

[2] Department of Physical Electronics, School of Electrical Engineering, Faculty of Engineering, Tel Aviv University, Tel Aviv 69978, Israel

[3] Laboratory of Nonlinear-Optical Informatics, ITMO University, St. Petersburg 197101, Russia

*Corresponding author:soumendujana@yahoo.com

We study the generation of dissipative solitons (DSs) in the model of the fiber-laser cavities under the combined action of cubic-quintic nonlinearity, multiphoton absorption and/or multiphoton emission (nonlinear gain) and gain dispersion. A random component of the group-velocity dispersion (GVD) is included too. The DS creation and propagation is studied by means of a variational approximation and direct simulations, which are found to be in reasonable agreement. With a proper choice of the gain, robust DS operation regimes are predicted for different combinations of multiphoton absorption and emission, in spite of the presence of the perturbation in the form of the random GVD. Importantly, the zero background around the solitons remains stable in the presence of the (necessary) linear gain. The solitons are stable too against a certain (realistic) level of noise. Another essential finding is that the quintic gain in the form of three-photon emission (3PE) offers an alternative mechanism for supporting stable solitons, provided that it is not too strong. The DSs coexist in low- and high-amplitude forms, for a given value of their width. The low-amplitude DS is stable, while its high-amplitude counterpart is subject to the blowup instability, in the presence of the 3PE. Interactions between DSs show various scenarios of the creation of breather states through merger of the two solitons.

## 1. INRODUCTION

Fiber lasers draw steadily growing interest due to their high output power, compact size, low manufacturing cost, and reliable performance, including robust mode-locking operation [1-4]. A wide range of applications, from cutting and welding to telecommunications, rely upon the use of these lasers. As a result, the growth rate of manufacturing fiber lasers (14% in 2014 and 13% in 2015) far outstrips that of solid-state (-3% in 2014 and 2015) and carbon-dioxide lasers (2% in 2014 and -1% in 2015) [5]. Initially, some performance parameters (such as the pulse power, duration etc.) of fiber lasers were inferior to those of solid-state sources. But improvement of the design has made fiber lasers strong competitors of their solid-state counterparts.

Laser cavities are essentially dissipative systems. For this reason, the operation of fiber lasers in the pulsed regime can be efficiently modelled in terms of dissipative solitons (DSs), which are stable localized modes formed in nonlinear dispersive media featuring the interplay of gain and loss. Being created in a lossy medium, DSs need continuous supply of energy for their existence [6-9]. Generally, DSs arise as the result of two coupled balances. One is the equilibrium between diffraction (or group-velocity dispersion, GVD) and nonlinear self-focusing, as in any soliton-bearing system, including conservative ones. The other condition is the balance between gain and loss in the system, which is a requirement specific to dissipative systems. In contrast to conservative [10] and PT-symmetric [11,12] models, in which solitons exist in continuous families, parameters of DSs, such as their width, amplitude, velocity, etc., are uniquely selected by the two balances, and do not depend on initial conditions, in agreement with the fact that stable DSs are *attractors* in the dynamics of the dissipative system [13]. These two balance conditions hold in fiber lasers operating in the DS regime.

The stability of passive [14-15] and active [16] mode-locking regimes of the operation of fiber lasers helps to make them compact sturdy devices operating in the alignment-free mode. Passive mode-locked fiber lasers are also popular as sources of ultra-short pulses [17]. The inclusion of a saturable absorber (SA) into the fiber-laser circuit helps to achieve the passive mode-locking. In particular, a semiconductor saturable absorber mirrors (SESAM) may operate as an SA [1,2]. Recently, carbon nanotubes and graphene were also used as SAs [3, 4]. Besides that, artificial SA can be realized by employing a variety of other techniques, such the use of nonlinear couplers, second-harmonic-generating elements. Nonlinear polarization rotation [18-20] and nonlinear polarization-evolution saturable absorbers [21, 22] are widely used in fiber lasers, especially when high-performance is desired. In addition to commonly used single-wavelength fiber lasers [17, 23], multi-wavelength operation has been reported too [24]. Another variety, *viz.*, fiber disk lasers are getting quick acceptance due to their customized applications [25].

Interaction of two DSs in a fiber laser may lead to the formation of a stable bound state with the group velocity different from that of a single soliton [26-31]. Further, collisions between bunched complexes of two or several DSs and a free one may lead to the absorption of the incident free soliton by the bound state [32, 33].

Based on the spectral filtering of highly chirped pulses, a new kind of high-energy femtosecond modes has been observed in an Ytterbium-doped fiber laser in the all-normal-GVD configuration [34]. Similar settings have been used to generate DSs in normal-GVD fiber lasers [35-37], in which the DS energy strongly increases, while the pulse becomes highly chirped (unchirped bright solitons cannot exist under the normal GVD), and its spectrum attains an approximately rectangular shape with two or three local maxima [38]. Optical materials may feature saturable nonlinearity, which is often approximated by the cubic-quintic (CQ) form, i.e., a combination of self-focusing cubic and self-defocusing quintic terms, which correspond to real parts of the cubic ($\chi^3$) and quintic ($\chi^5$) susceptibilities. On the other hand, imaginary parts of $\chi^3$ and $\chi^5$ represent the two-photon absorption (TPA) and three-photon absorption (3PA) effects, respectively. These higher order nonlinear effects may play an important role in fiber-laser cavities.

Theoretical models mostly refer to an ideal fiber, which has a constant core diameter with fixed material parameters and doping density. However, real fibers may feature various imperfections, including shape variations, inhomogeneities of the refractive index, fluctuation in the dopant concentration, and effects of bending and ellipticity due to external stress. Random fluctuations of the fiber's core diameter (that lies in the range of $\pm(3$ to $4)\%$ for real fibers) and other imperfections give rise to random variations of the GVD. Even weak GVD fluctuations produce significant cumulative effects in the course of long transmission in fiber systems**.** The influence of the random GVD becomes more prominent for shorter pulses, especially for femtosecond ones, being the major cause of bit-pattern destruction in ultra-short pulses [39-42]. Moreover, random fiber lasers have been developed (with the randomness provided by disordered distribution of doping nanoparticles) which operate in a specific diffusive regime, which resembles one affected by the random GVD, and provide for very high efficiency [43-45].

Although DSs were vastly studied, considering combination of two or three of the above-mentioned ingredients (e.g., SA, TPA, gain dispersion, etc.), the analysis still needs to be carried out for a full set of the higher order-nonlinear effects, as well as including the randomness emulating the situation in realistic fiber-laser systems. For example, while effects of the TPA and gain dispersion on the pulse propagation have been studied [46, 47], formation of solitons in the presence of random GVD and multi-photon absorption has not been considered in detail. In the present work, we address the propagation of pulses and subsequent formation of DSs in a fiber laser under effects of the random GVD, CQ nonlinearities, and multiphoton absorption (i.e., TPA and 3PA). We also include the gain dispersion, as it plays a prominent role for ultra-short pulses propagating through an active gain medium. Due to the presence of the CQ nonlinearity, the solitons are expected to be bistable. An appropriate model of such dissipative systems is provided by the complex Ginzburg-Landau equation (CGLE) with CQ terms [48, 49].

The paper is organized as follows. The model is introduced in Section 2. It is considered by means of the variational approximation (VA) in an analytical form, with the objective to derive pulse-evolution equations. In Section 3, the generation of DSs in the fiber-laser model is addressed. In particular, approximate VA results are compared with numerical ones. Interactions of the DSs in the fiber laser are considered in Section 4, which is followed by a conclusion in Section 5.

## 2. THE MODEL

The propagation of ultra-short pulses in a lossy dispersive fiber with the CQ nonlinearity, gain dispersion, TPA, 3PA, and randomly varying GVD is governed by the following version of the CGLE [26, 32, 34]:

$$i\frac{\partial E}{\partial z} + \frac{D(z)}{2}\frac{\partial^2 E}{\partial t^2} + |E|^2 E - \gamma |E|^4 E$$
$$= \frac{i}{2}(g_o - \alpha)E + \frac{id}{2}\frac{\partial^2 E}{\partial t^2} - iK|E|^2 E - i\nu |E|^4 E. \quad \textbf{(1)}$$

Here $E$, $z$, and $t$ are the amplitude of the electromagnetic wave, propagation distance, and retarded time, respectively. GVD coefficient $D(z)$ in Eq. (1) includes random variations added to the constant GVD. The third and fourth terms represent cubic and quintic nonlinearities, respectively. The original cubic and quintic nonlinear coefficients are $n_2 = 3\text{Re}(\chi^{(3)})/8n_0$ and $n_4 = 5\text{Re}(\chi^{(5)})/16n_0$, where $n_0$ is the linear refractive index. Equation (1) is scaled so as to make the effective cubic coefficient equal to 1, while $\gamma$ is respective quintic coefficient, proportional to $n_4/n_2$. As we choose the self-focusing cubic and defocusing quintic nonlinearities, which provides for the stabilization of solitons [50, 51], $\gamma$ is positive. If the pulse's temporal width is larger than the intra-band relaxation time, the gain spectrum, $g(\omega)$, can be expanded in the Taylor series about the carrier frequency $\omega_0$. This leads to the first two terms on the right-hand side of Eq. (1), where $g_o$ and $d$ are the gain saturation and gain dispersion coefficients, respectively, while $\alpha$ stands for the dimensionless loss coefficient. The imaginary part of $\chi^{(3)}$ gives rise to the TPA coefficient, $\alpha_2 = 3\omega \, \text{Im}(\chi^{(3)})/2n_0^2 c^2 \varepsilon_0$, where $\varepsilon_0$ is the vacuum permittivity [52]. Likewise, the imaginary part of the fifth-order susceptibility $(\chi^{(5)})$ gives rise to the 3PA coefficient, $\alpha_3 = 5\omega \, \text{Im}(\chi^{(5)})/2n_0^3 c^3 \varepsilon_0^2$. In Eq. (1), $\alpha_2$ and $\alpha_3$ are scaled to be to $K$ and $\nu$.

It is usually expected that higher-order effects in fiber lasers are represented by the higher-order GVD and Raman scattering. Here, we chiefly consider the setting in which both these effects are absent. Indeed, in mode-locked fiber lasers, the higher-order dispersion can be compensated [53, 54] using segments of specialty fibers (e.g., photonic crystal fibers, or multimode ones in which a higher-order mode is used), or with the help of chirped fiber Bragg gratings, as well as bulk components, such as pairs of diffraction gratings. The Raman scattering, whose effect on high-power may be detrimental, can be suppressed too, using special fiber designs [55], or long-period gratings [56], or chirped-pulse amplification, or, also, hollow-core photonic-crystal fibers filled by an inert gas [57, 58].

To consider the governing equation (1) in an analytical form, we use the VA in conjugation with the Rayleigh's dissipation function (RDF), which has been widely used to describe the nonlinear pulse propagation in dissipative optical fibers [59, 60]. In the framework of this method, the CGLE is separated into conservative and dissipative parts, which correspond to the left- and right-hand sides of Eq. (1), respectively. The Lagrangian density for the conservative part of is

$$L = \frac{i}{2}(EE_z^* - E^*E_z) + \frac{D}{2}|E_t|^2 - \frac{1}{2}|E|^4 + \frac{\gamma}{3}|E|^6, \quad (2)$$

The RDF density, which takes care of the dissipative terms, can be constructed as

$$R = -\frac{i}{2}(g_o - \alpha)(EE_z^* - E^*E_z) + \frac{id}{2}(E_{tt}^*E_z - E_{tt}E_z^*) + iK|E|^2(EE_z^* - E^*E_z) + i\nu|E|^4(EE_z^* - E^*E_z), \quad (3)$$

We adopt the usual sech ansatz for a bright soliton of Eq. (1):

$$E(z,t) = A(z)\text{sech}\left(\frac{t}{W(z)}\right)\exp(i\phi(z)), \quad (4)$$

where $A(z)$, $W(z)$ and $\phi(z)$ represent the complex amplitude, temporal pulse width and phase, respectively. The total Lagrangian and RDF can be found by inserting ansatz (4) in Eqs. (2) and (3):
$\mathcal{L} = \int_{-\infty}^{\infty} L dt$ and $\mathcal{R} = \int_{-\infty}^{\infty} R dt$. The calculation yields

$$\mathcal{L} = i(AA_z^* - A^*A_z)W(z) + 2W(z)|A(z)|^2\frac{\partial \phi(z)}{\partial z} + \frac{D}{3}\frac{|A(z)|^2}{W(z)} - \frac{2}{3}|A(z)|^4 W(z) + \frac{16}{45}\gamma|A(z)|^6 W(z), \quad (5)$$

$$\mathcal{R} = -(g_o - \alpha)W(z)\left[2|A|^2\frac{\partial\phi(z)}{\partial z} + i(AA_z^* - A^*A_z)\right] + \frac{id}{3W(z)}\left[(AA_z^* - A^*A_z) - 2i|A(z)|^2\frac{\partial\phi(z)}{\partial z}\right] + \frac{4i}{3}KW(z)|A(z)|^2(AA_z^* - A^*A_z) + \frac{8}{3}KW(z)|A(z)|^4\frac{\partial\phi(z)}{\partial z} + \frac{32}{15}\nu W(z)|A(z)|^6\frac{\partial\phi(z)}{\partial z} + \frac{16i}{15}\nu W(z)|A(z)|^4(AA_z^* - A^*A_z), \quad (6)$$

Using the Euler-Lagrange equations,

$$\frac{d}{dt}\left(\frac{\partial \mathcal{L}}{\partial \dot{q}_j}\right) - \frac{\partial \mathcal{L}}{\partial q_j} + \frac{\partial \mathcal{R}}{\partial \dot{q}_j} = 0, \quad (7)$$

where, $\dot{q}_j = dq_j/dt$, $q_j$ being parameters $A(z)$, $A^*(z)$, $W(z)$ and $\phi(z)$, the following four equations are obtained:

$$-i\frac{d}{dz}(W(z)A^*(z)) = iW(z)A_z^*(z) + 2W(z)A^*(z)\frac{\partial\phi}{\partial z} + \frac{D}{3}\frac{A^*(z)}{W(z)} - \frac{4}{3}|A(z)|^2A^*(z)W(z) + \frac{id}{3}\frac{A^*(z)}{W(z)} + \frac{4i}{3}K|A(z)|^2A^*(z)W(z) + \frac{16}{15}\gamma|A(z)|^4A^*(z)W(z) + \frac{16}{15}i\nu|A(z)|^4A^*(z)W(z) - iW(z)(g_o - \alpha)A^*(z), \quad (8)$$

$$i\frac{d}{dz}(W(z)A(z)) = -iW(z)A_z(z) + 2W(z)A(z)\frac{\partial\phi}{\partial z} + \frac{D}{3}\frac{A(z)}{W(z)} - \frac{4}{3}|A(z)|^2A(z)W(z) - \frac{id}{3}\frac{A(z)}{W(z)} - \frac{4i}{3}K|A(z)|^2A(z)W(z) + \frac{16}{15}\gamma|A(z)|^4A(z)W(z) - \frac{16}{15}i\nu|A(z)|^4A(z)W(z) + iW(z)(g_o - \alpha)A(z), \quad (9)$$

$$i(AA_z^* - A^*A_z) = -2|A(z)|^2\frac{\partial\phi(z)}{\partial z} + \frac{D}{3}\frac{|A(z)|^2}{W^2(z)} + \frac{2}{3}|A(z)|^4 - \frac{16}{45}\gamma|A(z)|^6, \quad (10)$$

$$2W(z)\left[A(z)\frac{\partial A^*(z)}{\partial z} + \frac{\partial A(z)}{\partial z}A^*(z)\right] + 2|A(z)|^2\frac{\partial W}{\partial z} + \frac{2d|A(z)|^2}{3W(z)} + \frac{8K}{3}|A(z)|^4 W(z) + \frac{32\nu}{15}|A(z)|^6 W(z) - 2(g_o - \alpha)|A|^2 W(z) = 0, \quad (11)$$

Further manipulations with the equations yield balance relations for the pulse propagation. Namely, multiplying Eq. (8) and Eq. (9) by $A(z)$ and $A^*(z)$, respectively, and subsequent subtraction yields

$$\frac{d}{dz}(2|A(z)|^2 W(z)) = -\frac{2d|A(z)|^2}{3W(z)} - \frac{8}{3}KW(z)|A(z)|^4 - \frac{32}{15}\nu W(z)|A(z)|^6 + 2W(z)(g_o - \alpha)|A|^2, \quad (12)$$

The integrated intensity of the pulse is $\int_{-\infty}^{+\infty}|E(z,t)|^2 dt = 2|A(z)|^2 W(z)$. Therefore, Eq. (12) describes the variation of pulse's energy in the course of the propagation.

Further, multiplying Eq. (8) and (9) by $A(z)$ and $A^*(z)$, respectively, and adding the results, one obtains

$$i(AA_z^* - A^*A_z) = -2|A(z)|^2\frac{\partial\phi(z)}{\partial z} - \frac{D}{3}\frac{|A(z)|^2}{W^2(z)} + \frac{4}{3}|A(z)|^4 - \frac{16}{15}\gamma|A(z)|^6, \quad (13)$$

Comparing Eqs. (10) and (13), we get

$$\frac{|A(z)|^2 W^2(z)}{D(z)} - \frac{16\gamma|A(z)|^4 W^2(z)}{15 D(z)} = 1 \quad (14)$$

Equation (14) is the fundamental constraint for the pulse propagation. Notably, no contribution from the dissipative part appears in Eq. (14). To have a solution obeying this condition, one should ensure the gain-loss balance.

Equations (12) and (14) give rise to the evolution equations for the pulse's amplitude, $A(z)$ and width, $W(z)$:

$$\frac{dW(z)}{dz} = \left[\frac{-d}{3sW}(1-\sqrt{M}) - \frac{2KW}{3s^2}(1-\sqrt{M})^2 \right.$$
$$\left. - \frac{4vW}{15s^3}(1-\sqrt{M})^3 - \frac{2}{W\sqrt{M}}\frac{dD}{dz} \right.$$
$$\left. + W(z)(g_o - \alpha)\frac{1}{s}(1-\sqrt{M})\right]$$
$$/\left[\frac{-4D}{W^2\sqrt{M}} + \frac{1}{s}(1-\sqrt{M})\right], \quad (15)$$

$$\frac{dA(z)}{dz} = \left[\frac{1}{2AW^2 - 4sW^2A^3}\right]\frac{dD}{dz} - A(1-sA^2)\left[\frac{-d}{3sW}(1-\sqrt{M})\right.$$
$$\left. - \frac{2KW}{3s^2}(1-\sqrt{M})^2 - \frac{4vW}{15s^3}(1-\sqrt{M})^3 - \frac{2}{W\sqrt{M}}\frac{dD}{dz} \right.$$
$$\left. + W(z)(g_o - \alpha)\frac{1}{s}(1-\sqrt{M})\right]$$
$$/\left(W(1-2sA^2)\left[\frac{-4D}{W^2\sqrt{M}} + \frac{1}{s}(1-\sqrt{M})\right]\right), \quad (16)$$

where $M = 1 - 4sD/W^2$ and $s = 16\gamma/15$.

## 3. THE DISSIPATIVE PULSE DYNAMICS AND GENERATION OF DISSIPATIVE SOLITONS

The evolution of the pulse's parameters is predicted by solving Eqs. (15) and (16). Subsequently DSs can be generated in the approximate form, using these solutions for a suitable set of parameters. The investigation is done under two the conditions of (i) the balance between the nondispersive gain and linear loss, $g_o = \alpha$ or (ii) in the absence of the balance: $g_o \neq \alpha$.

**CASE (I)** $g_o = \alpha$ **(THE NONDISPERSIVE-GAIN – LINEAR-LOSS BALANCE)**

To examine the influence of the nonlinear losses (i.e., TPA and 3PA) and the gain dispersion, we adopt condition $g_o = \alpha$, to eliminate the linear gain and loss in the system. Overall, the system still remains dissipative under the effect of the TPA, 3PA and gain-dispersion terms. The GVD coefficient, $D(z)$, comprises a constant part (scaled to be 1) plus a randomly varying part ($\epsilon$) with mean value 0.03. The effect of the finite gain bandwidth must be taken into account because the femtosecond-pulses' spectra are very wide [61]. The gain dispersion in conjugation with the GVD significantly affects the pulse dynamics and energy profile.

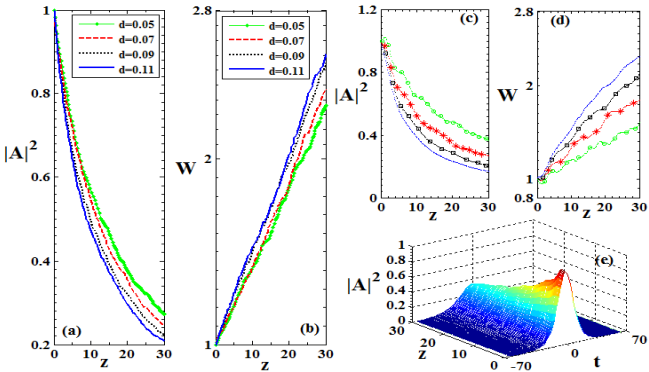

**Fig. 1**. The pulse degradation under the combined effect of the TPA, 3PA, gain dispersion, and random GVD, while the nondispersive gain is in balance with the linear loss. (a) The decay of pulse's intensity, and (b) the increase of its width for different values of gain dispersion $d$, as produced by the solution of variational equations (15) and (16). Here, $K$=0.01, $v$= 0.01, $\gamma$= 0.1. Counterparts of these results, as obtained from direct simulations of Eq. (1) with $d$=0.05 (green line with circles), $d$=0.07 (red line with stars) $d$=0.09 (black line with squares) $d$=0.11 (blue line) are displayed in panels (c) and (d), respectively. (e) The numerically generated 3D profile of the dissipative-pulse's evolution for $d$=0.05.

Variational results shows that the gain dispersion reduces its intensity (Fig. 1(a)) and makes it broader (Fig. 1(b)), causing degradation of the pulse quality.

The verification of the variational results is provided by direct simulations of Eq. (1), using the split-step Fourier method, see Figs. 1(c-e). Here we set the integration stepsize equal to $10^{-3}$. The propagation length is taken to be tantamount to 30 soliton periods.

The systematic pulse decay, observed in Fig. 1(c), and its broadening, as seen in Fig. 1(d), are close to the averaged VA-predicted counterparts. The fluctuations in the VA curves and ones generated by the full simulations are different, representing an effect of the random part of the GVD.

Being the third-order nonlinear-loss phenomenon, the TPA has a quadratic dependence on the amplitude of the electromagnetic wave. Generally, it limits the efficiency of optical switching and causes reshaping and broadening of solitons, as well as splitting of higher-order ones into constituents fundamental pulses [62]. In the general case, the number of the emerging solitons depends on the gain and length of the fiber amplifier, if it is present in the system [63, 64]. The TPA coefficient is estimated as $6.2 \times 10^{-15}$mW$^{-1}$ in As$_2$S$_3$-based glass at 1.55 μm [65]. The 3PA leads to a higher degree of confinement, as it is proportional to the fourth power of the field amplitude, and has potential applications to wavelength shifting, pulse reshaping, and stabilization in narrow-pulse fiber communication systems [66]. In As$_2$S$_3$-based glass, the 3PA coefficient is $2.0 \times 10^{-27}$ m$^3$W$^{-2}$ at the wavelength of 1.55 μm [67]. Recently, both saturable absorption and TPA of few-layer molybdenum diselenide (MoSe$_2$) have been observed at 1.56 μm wavelength, and subsequently used in an all-fiber Erbium-doped mode-locked ultrafast fiber laser [68].

Currently, much attention is drawn to mid-infrared (IR) wavelengths (> 3000 nm), and the operation of fiber lasers based on materials which are appropriate in this range, such as chalcogenide glasses. With this in mind, throughout this work we adopt physical parameters relevant to the mid-IR wavelengths, where the higher-order nonlinear effects are prominent [69]. Actually, all analysis throughout this paper is performed for parameters corresponding to wavelength 3500 nm propagating in chalcogenide fibers.

The nonlinear absorption in the fiber reduces the pulse's intensity. It can also represent injection of electron-hole pairs, leading to the free-carrier absorption and dispersion. TPA and 3PA individually have detrimental effect on the pulse propagation, the decay rate caused by 3PA being smaller than its TPA-induced counterpart. Actually, the TPA is more prominent at shorter wavelengths, while 3PA is dominant at longer wavelengths, both effects being essential in the intermediate region [65].

The pulse's dynamics is more interesting in the presence of a negative imaginary part of the $\chi^{(5)}$ coefficient, i.e., a negative 3PA coefficient, which implies the quintic gain, rather than loss. The resulting amplification effect is sometimes called the three-photon emission (3PE) [70]. In particular, the action of the polarization-correlated 3PE leads to formation of a positively charged *triexciton* (a bound state of three electron-hole pairs) in a self-assembled GaAs quantum dot [71].

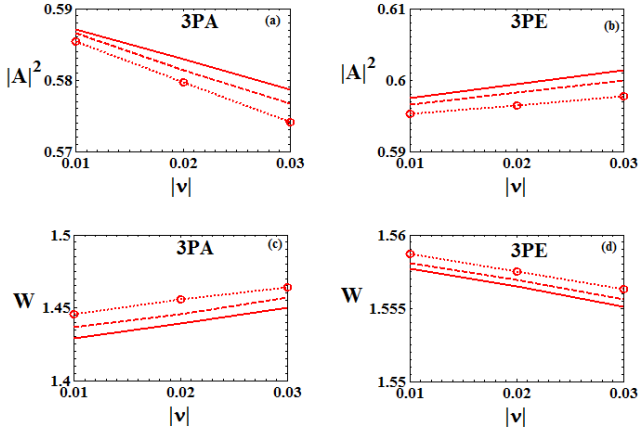

**Fig. 2**. Variation of (a) the normalized pulse's intensity and (c) its width with the increase of $|\nu|$ in the cases of the 3PA. (b) and (d) are the same for 3PE. In all the panels, solid lines correspond to $\gamma$=0.03, while dashed and dotted lines correspond to $\gamma$= 0.05 and 0.1 respectively. Other parameters are $K$= 0, and $d$=0.05.

In semiconductor devices, the cubic gain, i.e., two-photon emission, was observed in optically pumped GaAs and in current-driven GAInP/ AlGaInP quantum wells [72]. However, experimental realization of the 3PE still has to be elaborated. Our VA results show that, with the increase of $|\nu|$, while the 3PA leads to the decay of the pulse's intensity (Fig. 2(a)) and its broadening (Fig. 2(b)), 3PE naturally causes the opposite, i.e., growth of the intensity and decrease of the pulse's width. We stress that the 3PE does not lead to blow-up of the pulses, as the 3PE coefficient considered here is small enough.

## CASE (II) $g_o > \alpha$ (THE NONDISPERIVE GAIN GREATER THAN THE LINEAR LOSS)

The pulse degradation discussed above can be arrested, and a DS can be built, by applying proper gain to the system, which makes $\Delta g \equiv g_0 - \alpha > 0$ in Eq. (1). Strictly speaking, this condition, i.e., the presence of the *excess linear gain*, makes the zero background unstable around any soliton. Nevertheless, it is shown below that the background instability may be avoided, in a properly chosen setting (in particular, limiting the excess gain to sufficiently small values, $\Delta g \sim 10^{-5}$, see below). This may be explained by the fact the small disturbances are set in motion by the GVD, which then adds effective loss due to the gain dispersion, and, eventually, the disturbances hit edges of the integration domain, or the DS; in the latter case, the nonlinear dissipation can help to suppress them. Furthermore, the presence of CQ terms in the system suggests (see, in particular, Eq. (14) that DSs can be made bistable. Figure 3(a) shows the bistability: for a fixed pulse's width (e.g.,$W$ = 1.25), two amplitudes ($A$ = 0.853 and $A$ = 2.947) are obtained from the curve corresponding to $\gamma$ = 0.1. This means that a 12 ps pulse can generate dissipative solitons of power 87 mW and 750 mW.

Direct numerical simulations reveal that the smaller-amplitude DS is stable for $\Delta g = 4.35 \times 10^{-6}$, see Fig. 3b (i), while the larger-amplitude one eventually blows up, in Fig. 3b(ii), at the same gain. However, the latter DS may be made quasi-stable by choosing suitable loss, namely, $\Delta g = -1.69 \times 10^{-5}$, see Fig. 3b (iii). Here, $\gamma$ = 0.1, $W$ = 1.25 and $A$ = 0.853 or 2.947 are chosen for the explicit presentation of the results. Similar results are obtained by choosing other sets of values of $\gamma$, $W$ and $A$ from the bistability curves in Fig. 3(a).

It is known that solitons in complex models may feature internal modes, which manifest themselves as persistent oscillations of the soliton's shape [73]. In our system, small-amplitude shape oscillations are observed in stable solitons with the lower amplitude, see Fig. 3b (i).

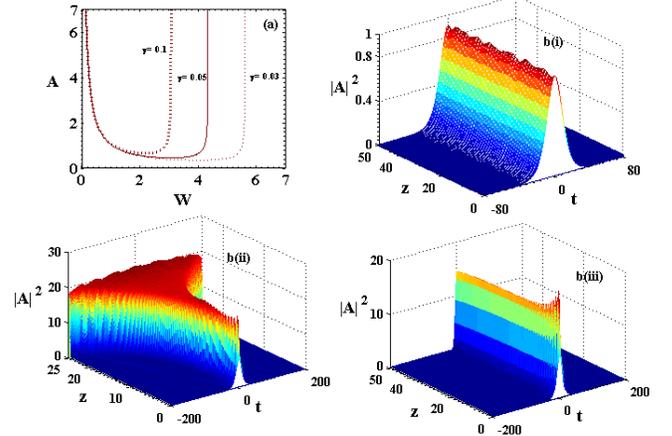

**Fig. 3**. (a) The bistability curve corresponding to Eq. (14) for $\gamma$ = 0.1, 0.05 and 0.03. (b) Numerically generated pulse evolution pertaining to the same width ($W$ =1.25) but different amplitudes selected from (a) for $\gamma$ = 0.1. In panel (b), (i) displays the evolution of a stable DS corresponding to the smaller amplitude ($A$ = 0.853) for excess gain $\Delta g = 4.35 \times 10^{-6}$, while (ii) shows the blowup of an unstable DS with the larger amplitude ($A$ = 2.947) and the same excess gain. In addition, panel b (iii) displays quasi-stabilization of the higher-amplitude DS with $A$ = 2.947, in the presence of a very weak effective loss, viz., $\Delta g = -1.69 \times 10^{-5}$. In panels (b), the parameters are $K$= 0.01, $\nu$= 0.01.

The present study does not include the Raman gain. It can be directly verified that the inclusion of the Raman term with realistic values of parameters into the present model does not destabilize the dissipative solitons, as shown in Figure 4.

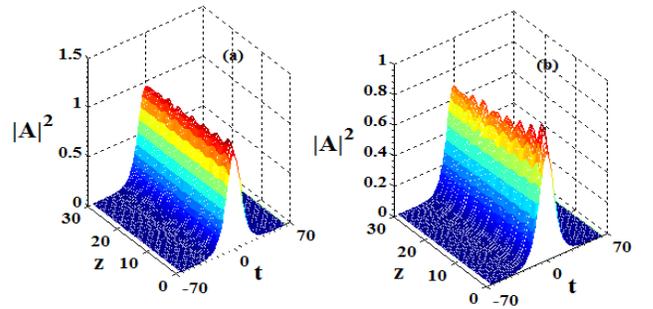

Fig. 4: The generation of a stable dissipative soliton in the presence of the Raman gain with different gain coefficients: (a) 0.01, (b) 0.001. $\Delta g = 4.34 \times 10^{-6}$ and $4.50 \times 10^{-6}$ for (a) and (b) respectively. For both the panels, $\gamma$ = 0.1, K=0.01, d = 0.05 and $\nu$ = 0.01.

We now proceed to the analysis of the DS evolution in the presence of the TPA and 3PA/3PE terms. In the presence of the TPA, small excess gain, $\Delta g = 4.23 \times 10^{-6}$, results in a slightly fluctuating but generally steady peak intensity and pulse's width, as shown in Fig. 5. The fluctuations are more prominent in the full simulations, but mean values of the peak intensity and width almost exactly match their numerically computed counterparts ($< A^2 >_{num} = 1.0537$ and $< W >_{num} = 0.9727$, respectively).

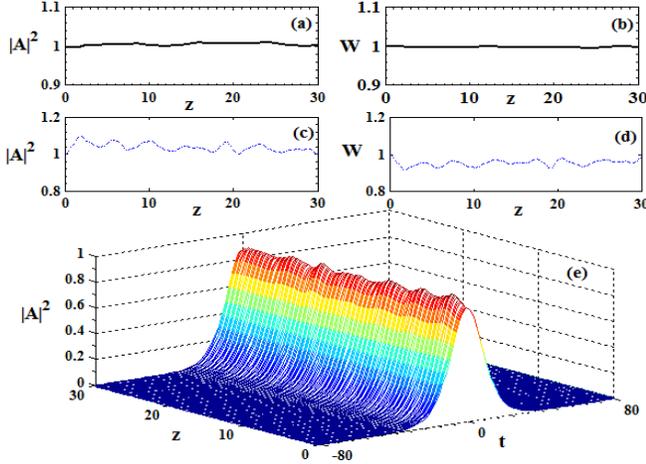

**Fig. 5.** The evolution of the DS in the presence of the TPA effect, while the 3PA term is absent. The corresponding variation of the soliton's intensity and width, as predicted by the VA, are shown in (a) and (b), respectively. Panels (c) and (d) display the same, but as obtained from direct simulations. The 3D plot of the numerically simulated evolution of the DS is displayed in (e). In all panels, γ = 0.1, $d = 0.05$ and $K = 0.01$, ν=0. Here the excess gain is $\Delta g = 4.23 \times 10^{-6}$

Further, the evolution of the DS under the action of the 3PA is shown in Fig. 6 with excess gain $\Delta g = 4.10 \times 10^{-6}$, which is somewhat less than that required to compensate the nonlinear loss in the case of the TPA. Mean values of the simulated peak intensity and width are $< A^2 >_{num} = 1.0247$ and $< W >_{num} = 0.9825$, that are close to variationally obtained values.

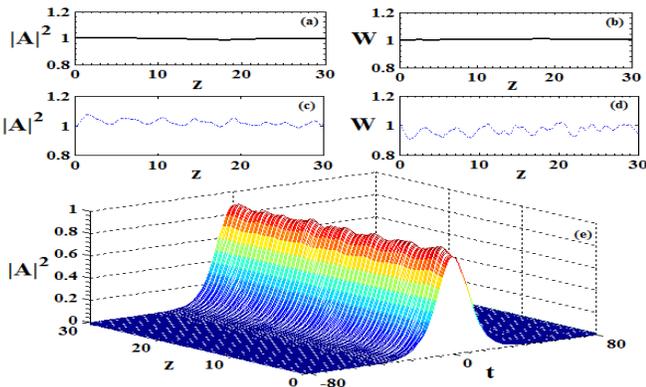

**Fig. 6**. The same as in Fig. 5, but for the case of the evolution of the DS under the action of the 3PA, in the absence of the TPA. Here, γ = 0.1, K= 0, $d = 0.05$ and ν = 0.01. Here excess gain $\Delta g = 4.10 \times 10^{-6}$

Naturally, a stronger excess gain is required (actually, it is $\Delta g = 4.35 \times 10^{-6}$) to form the DS in the presence of both the TPA and 3PA, see Fig. 7. In this case, mean values of the simulated peak intensity and width are $< A^2 >_{num} = 1.0298$ and $< W >_{num} = 0.9825$, respectively.

Thus, we conclude that the VA-predicted results generally match findings produced by the direct simulations in Fig. 5-7, although the numerical results show more prominent oscillations in the pulse's peak intensity and width. Nevertheless, the mean values of the numerically obtained peak intensity and width are almost exactly fitted by the VA-predicted counterparts.

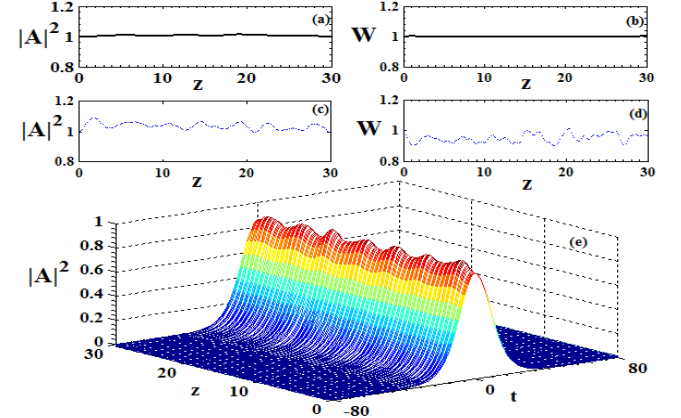

**Fig. 7.** The same as in Figs. 5 and 6, but in the case of the evolution of the DS under the combined action of the TPA and 3PA terms. In all panels, γ =0.1, K=0.01, $d = 0.05$ and ν = 0.01. Here excess the gain is $\Delta g = 4.35 \times 10^{-6}$.

Now, it is relevant to consider the 3PE as an alternate source of gain. Instead of the TPA-3PA combination considered above, the TPA-3PE one requires a smaller excess gain, $\Delta g = 4.15 \times 10^{-6}$ to form a DS, see Fig. 8. Thus, 3PE may indeed be harnessed as an alternative gain mechanism, provided that it is small enough to avoid the onset of the blowup. In that case, the use of the 3PE is actually a stabilizing factor, as it allows one to use a smaller linear excess gain, and thus improve the stability of the zero background. In the parameter region investigated here, the blowup is absent indeed for the DSs belonging to the left branch of the bistability curve in Fig. 3(a), while the solitons with the larger amplitude, belonging to the right branch, eventually do blow up.

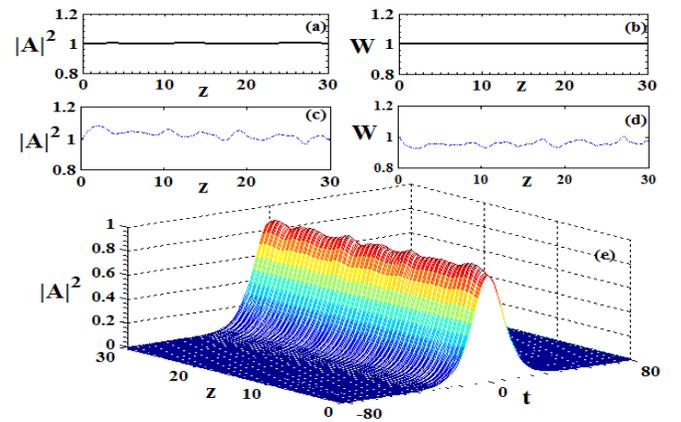

**Fig. 8**. The same as in Figs. 5 and 6, but for the evolution of the DS under the combined action of the TPA and 3PE terms. Here, γ = 0.1, K=0.01, $d = 0.05$ and ν = -0.01, while the excess gain is $\Delta g = 4.15 \times 10^{-6}$

In this connection, it is relevant to mention that stable states supported by the *unsaturated gain* (i.e., the higher-order amplification not capped by attenuation of a still higher degree) were previously found in some other models [74, 75]. A generic feature of such models is the existence of an unstable solution with a larger amplitude, which plays the role of a *separatrix*, i.e., a boundary between initial states which are attracted to a stable smaller-amplitude solution, and those which undergo the blowup. In the present case, the DSs belonging to the right branch of the curve in Fig. 3(a) plays this role.

Although the analysis presented in this section includes the random GVD, it is easy to check that this ingredient of the model is not responsible for the stability of the DSs, as they remain equally stable or unstable when solely the constant GVD is kept. On the other hand, the analysis of the system including this practically important term is relevant, as it additionally attests to the stability of the DSs in the presence of the random perturbations, see Figs. 3-8.

It is essential to take into account effects of random GVD on the soliton propagation. As the randomness enhances the bit-error-rate, its effect is obviously detrimental for the propagation of ultra-short pulses. Randomness of the GVD may be, in principle, both temporal and spatial. Possible temporal variation of the GVD being, in any case, much slower than the high-speed soliton pulse propagation, we here consider only the spatial randomness of the GVD, considering the local dispersion coefficient as a sum of a constant part and a randomly varying one ($\epsilon$). Typically, the mean value of $\epsilon$ is zero. To get an idea of robustness of the DS in the presence of the random inhomogeneity of the GVD, in Fig. 9 we display the pulse propagation for different magnitudes of the random GVD. Naturally, fluctuations of the pulse's intensity and width increase with the growth of the magnitude of the randomness. In the subsequent analysis, we fix the random-variation magnitude to be 3% of the constant part of the GVD coefficient. In this case, the analysis readily produces stable DSs, which are virtually identical to those displayed in Figs. 3(b), (d), and 5-8. In particular, the randomness at the 7% level still admits quite robust propagation of the pulse, as clearly seen in Fig. 10.

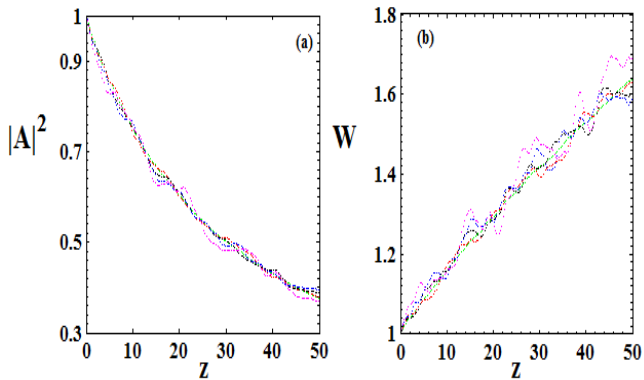

**Fig. 9**. (a) The pulse's intensity and (b) width for different relative magnitudes of the random part of the dispersion in comparison with its constant part, as obtained from direct simulations of Eq. (1). Color dots correspond to the following magnitudes: 0% (green), 3% (red), 5% (black), 7% (blue), and 9% (magenta). Here, γ = 0.1, K=0.01, d = 0.05 and ν = 0.01.

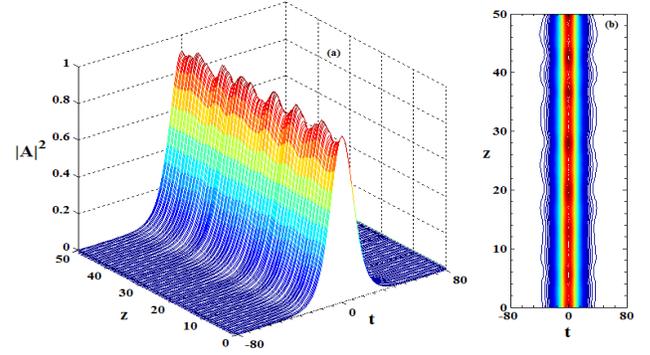

**Fig. 10**. (a) The propagation of a robust soliton in the presence of the random dispersion at the 7% level, with respect to its constant part. (b) The corresponding contour plot. Here γ = 0.1, K=0.01, d = 0.05 and ν = 0.01, and the excess gain is $\Delta g = 4.34 \times 10^{-6}$.

Along with the study of the effect of the random variation of the GVD coefficient along the fiber, it is necessary to address stability of the DSs against initial injunction of a random noise, a well-known source of which is the amplified spontaneous emission. Results of typical simulations of the noise-affected propagation of the DS are displayed in Fig. 11, which clearly demonstrate that the propagation remains robust even in the presence of a strong noise.

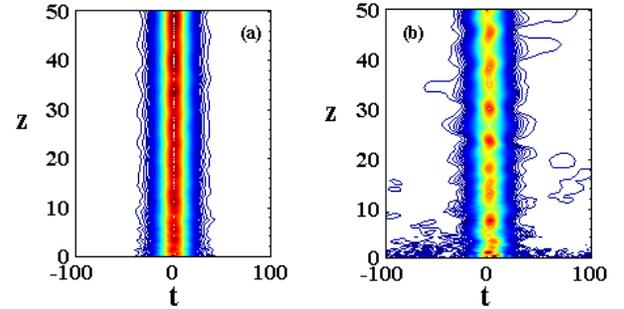

**Fig. 11**. Comparison of the evolution of the DS in the presence of an initial random noise at the 5% level (a), and 15% (b). Here γ = 0.1, K=0.01, d = 0.05 and ν = 0.01.

## 4. INTERACTIONS OF DISSIPATIVE SOLITONS IN THE FIBER LASER

One of the basic features of solitons in (nearly) integrable systems is that they preserve their identity upon collisions. Shifts produced by collisions open a possibility for "steering light by light" [52]. In this section, we study interactions between the DSs, which were constructed above, in two different ways: first, the interaction between in-phase solitons with different initial separations, and then the interaction between solitons with a constant initial separation but different relative phases, Δϕ. We have studied the interaction between two high-amplitude solitons (Fig. 12) as well as two low-amplitude ones (Fig. 13), originally separated by some distance (temporal delay). Initially, the DS pairs are taken to be in-phase, with zero relative velocity between them. The large-amplitude solitons eventually blow up. A suitable loss Δg is applied to make them quasi-stable. The ensuing interaction dynamics significantly depends on the initial separation. At smaller separations, the DSs exhibit periodic collision and eventually merge into a single DS (*breather*), that maintains periodic oscillation of its amplitude and width in the course of subsequent propagation. With the increase of the initial separation, the interaction becomes weaker, and ceases at the separation of $T_g \geq 50$.

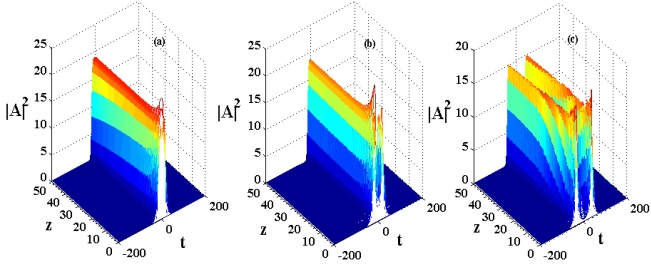

**Fig 12** The interactions between two in-phase high-amplitude ($A$ = 2.947) DSs for different initial temporal separations ($T_g$) between them: (a) $T_g$=10, (b) $T_g$=20, (c) $T_g$=35. Other parameters are γ= 0.1, d=0.05, K=0.01 and ν= 0.01. The linear loss for (a) $\Delta g = -3.80 \times 10^{-5}$ (b) $\Delta g = -3.80 \times 10^{-5}$ and for (c) $\Delta g = -1.7255 \times 10^{-5}$.

The interaction of small-amplitude DSs shows a different behavior. Instead of merging, they continue to coalesce and split periodically, thus exhibiting very robust breather-like propagation. The frequency of the periodic collisions decreases with the increase of the initial separation $T_g$ up to 65, beyond which the interaction ceases.

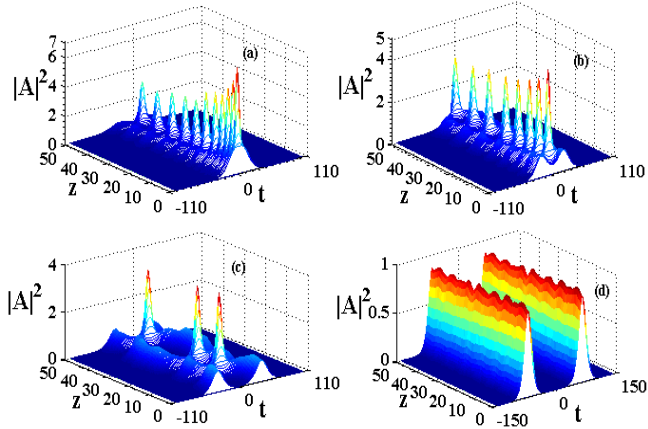

**Fig. 13.** The interactions between two in-phase low-amplitude ($A$ = 0.853) DSs for different initial separations ($T_g$) between them. (a) $T_g$=10, (b) $T_g$=20, (c) $T_g$=35, and (d) $T_g$=65. The corresponding top views are displayed in panels are (e), (f), (g) and (h), respectively. Other parameters are γ= 0.1, d=0.05, K=0.01 and ν= 0.01. The excess linear gain is $\Delta g = 4.35 \times 10^{-6}$.

More interesting phenomenology was observed, varying relative phase $\Delta\phi$ between two interacting solitons with a constant initial separation. Figure 14 portrays such interaction for initial separation $T_g$=10. In particular, for small $\Delta\phi = \pi/10$, one of the two interacting pulses quickly vanishes, transferring its energy to the other, which features deceleration in the course of subsequent propagation. At larger $\Delta\phi$, the energy transfer takes place quicker, and the deceleration decreases. At $\Delta\phi = \pi/2$, the two solitons undergo very fast merger, without any deceleration.

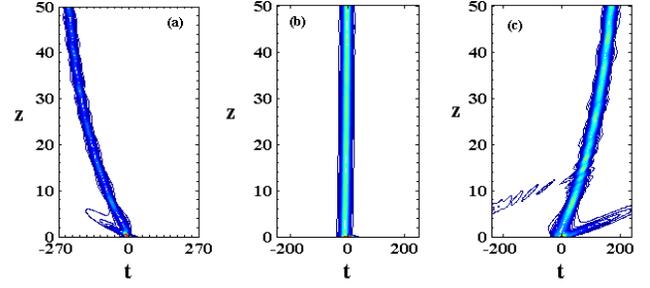

**Fig. 14** Relative-phase-controlled switching, featured by the interaction of two DSs at different relative phases but with a fixed initial separation $T_g$=10. The phase difference is $\Delta\phi = \pi/10$ in (a), $\pi/2$ in (b), and $\pi$ in (c). Other parameters are γ= 0.001 K=0.01, ν= 0.01, d= 0.01. Here the excess linear gain is $\Delta g = 4.32 \times 10^{-6}$.

Further increase of the relative phase gives rise to emission of radiation from one pulse and eventual transformation of the pair into a single pulse, which accelerates (on the contrary to the deceleration observed at $\Delta\phi < \pi/2$). The acceleration increases with the increasing of $\Delta\phi$ up to $\pi$. Figure 15 shows the temporal shift of the pulse as a function of $\Delta\phi$ at a fixed normalized propagation distance, $z$ = 30, for initial separation $T_g$=10.

The switching from the deceleration to acceleration stage occurs much faster at smaller $T_g$. For example, at $T_g$ = 10 the temporal-shift rate is 2.36/degree is observed, while at $T_g$ = 1 it is 3.32/degree.

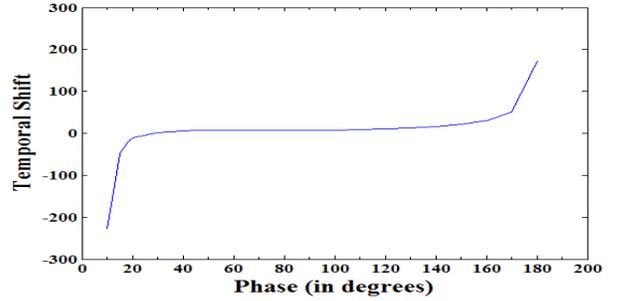

**Fig 15.** The temporal shift of the single DS emerging from the original pair (with initial $T_g$=10) versus the phase shift between the initial DSs. Other parameters are γ= 0.001, $K$= 0.01, ν= 0.01, $d$= 0.05.

## 5. CONCLUSION

We have studied the pulse propagation in the realistic model of fiber laser cavities under the action of the randomly varying GVD, loss, multiphoton absorption or emission (nonlinear gain), higher-order nonlinearity, and gain dispersion. We have found conditions for the stable operation of the laser in the DS (dissipative-soliton) regime. A nontrivial feature is the stability of the zero background around the solitons, in spite of the presence of the linear gain; an explanation of this feature was given. DSs were obtained in an approximate form by means of the VA (variational approximation) and direct simulations, with a conclusion that the quasi-analytical results produced by the VA are in reasonable agreement with the numerical findings. An essential result, produced by both methods, is that the nonlinear amplification, provided by the 3PE (three-photon emission, i.e., the quintic gain) provides for an efficient alternative gain mechanism for the stable DSs, provided that it is not too strong, to avoid the onset of the blowup. Another noteworthy fact is that the DSs remain stable under the action of the perturbation in the form of the random GVD as well as noise. The DSs are bistable, with two different pulses, low- and high-amplitude ones, found for a given width. In the presence of the nonlinear gain, the low-amplitude DS is stable, while its high-amplitude counterpart is subject to the blowup instability. Interactions between the DSs lead to fusion of high-amplitude solitons into breathers, and periodic merger-splitting sequences for low-amplitude ones. The results reported in the paper suggest new experiments for DSs in fiber lasers.

**Acknowledgements.** G. S. Parmar is grateful to TEQIP and Thapar University for providing financial support through teaching associateship.